\documentclass[aps,prb,twocolumn,floatfix,amsfonts,superscriptaddress,longbibliography]{revtex4-1}
\pdfpageattr {/Group << /S /Transparency /I true /CS /DeviceRGB>>}
\usepackage{amsmath}
\usepackage{amsthm}
\usepackage{amssymb}
\usepackage{graphicx}
\usepackage{tabularx}
\usepackage{color}

\newcommand{\cre}[2]{#1_{#2}^\dagger} 
\newcommand{\ann}[2]{#1_ {#2}^{\phantom{\dagger}}}

\begin{document}

\title{Validity of the local self-energy approximation: \\application to coupled quantum impurities}

\author{Andrew K. Mitchell}
\affiliation{Institute for Theoretical Physics, Utrecht University,
3584 CE Utrecht, The Netherlands}

\author{Ralf Bulla}
\affiliation{Institute for Theoretical Physics, University of Cologne,
  50937 Cologne, Germany}


\begin{abstract}
We examine the quality of the local self-energy approximation, applied here to models of multiple quantum impurities coupled to an electronic bath. The local self-energy is obtained by solving a single-impurity 
Anderson model in an effective medium that is determined self-consistently, similar to the dynamical mean-field theory (DMFT) for correlated lattice systems. By comparing to exact results obtained using the numerical renormalization group, we determine situations where ``impurity-DMFT'' is able to capture the physics of highly inhomogeneous systems, and those cases where it fails. For two magnetic impurities separated in real-space, the onset of the dilute limit is captured, but RKKY-dominated inter-impurity singlet formation cannot be described. For parallel quantum dot devices, impurity-DMFT succeeds in capturing underscreened Kondo physics by self-consistent generation of a critical pseudogapped effective medium. However, the quantum phase transition between high- and low-spin states on tuning interdot coupling cannot be described. 
\end{abstract}

\maketitle


\section{Introduction}

Dynamical mean field theory (DMFT) has been well-established as a powerful tool for treating correlated materials.\cite{georges1996dynamical} At its heart is a local self-energy approximation that only becomes exact in the limit of infinite lattice coordination.\cite{metzner1989correlated} In many applications to real materials,\cite{held2007electronic} direct comparisons with experiment have shown the quality of this approximation to be very good.  However, the interacting lattice fermion models to which DMFT is applied typically have no known exact solution with which to fully compare.

Although DMFT was originally formulated for translationally-invariant problems, the local self-energy approximation can also be used in an  inhomogeneous context. Each inequivalent site in the system is mapped to a single-impurity Anderson model in an effective medium that is determined self-consistently.\cite{potthoff1999surface,*potthoff1999metallic} This inhomogeneous (or real-space) DMFT has been applied to a diverse range of systems, including materials with surfaces\cite{potthoff1999surface,*potthoff1999metallic}, layered heterostructures,\cite{freericks2004dynamical,helmes2008kondo,okamoto2004spatial} atoms in optical traps,\cite{helmes2008mott,snoek2008antiferromagnetic} and even nanostructures or devices for molecular electronics.\cite{florens2007nanoscale,jacob2010dynamical,turkowski2012dynamical} 

Inhomogeneous DMFT has also been used to study depleted lattices.\cite{kaul2007strongly,dobrosavljevic2005absence} In the case of metals doped with randomly distributed magnetic impurities, the physics is dominated by effects of disorder; while heavy fermion physics is recovered on approaching the periodic limit, where standard DMFT applies. In the other limit of dilute impurities, inhomogeneous DMFT is exact by construction since the impurity effective medium is simply the clean host metal.

The physics of dilute magnetic impurities is of course also well-known,\cite{hewson1997kondo} with the impurity spin degrees of freedom being screened by surrounding conduction electrons by the Kondo effect at low temperatures.\cite{CoonCu_Nikolaus,*CoonCu_Manoharan,*wahl_dens} Models of single magnetic impurities can be solved exactly with the numerical renormalization group\cite{bulla2008numerical,costi2009kondo,*hanl2013iron,derry2015quasiparticle} (NRG), which provides access to thermodynamic and dynamic quantities on essentially any temperature or energy scale. 

However, the physics of the intermediate regime with a few coupled quantum impurities is notoriously complex. Already in the two-impurity case, there is a subtle competition between single-impurity Kondo physics and through-host RKKY coupling.\cite{jones1987study,*jones1988low,sakai1992excitation,silva1996particle,allerdt2015kondo} Theoretically, multi-impurity problems are inherently much more difficult to treat: $N$-impurity models generally involve $N$ coupled conduction electron channels. For NRG, the required computational resources scale exponentially with the number of channels. In fact, full dynamical properties of the two-impurity system have only recently been obtained with NRG for a true real-space system, taking into account details of the host lattice exactly.\cite{mitchell2015multiple} 
Although the efficiency of NRG can be greatly enhanced by interleaving different Wilson chains\cite{mitchell2014generalized,iNRG} or exploiting large symmetries,\cite{weichselbaum2012non} it is clear that an exact solution for true many-impurity systems is beyond reach. Similar difficulties afflict other numerical methods.

Coupled quantum dot devices\cite{pustilnik2004kondo} and magnetic nanostructures or single-molecule junctions\cite{florens2007nanoscale,jacob2010dynamical,turkowski2012dynamical}  can also be described theoretically in terms of generalized quantum impurity problems. However, these systems typically contain many correlated `impurity' degrees of freedom (and several effective screening channels), again essentially precluding an exact numerical treatment.

Inhomogeneous DMFT is therefore a highly appealing approximate technique for such systems, since the computational complexity grows only linearly in the number of inequivalent interacting sites.\cite{potthoff1999surface,*potthoff1999metallic} However, details of geometry (or molecular structure), inter-impurity coupling and RKKY effects are important in these contexts. It is therefore not at all obvious what the quality of the local self-energy approximation is, since these systems are  far from an infinite-coordination limit.

In this paper we apply inhomogeneous DMFT
to two paradigmatic quantum impurity problems --- two magnetic
impurities separated in real-space, and systems of coupled quantum dots  between metallic leads (see Fig.~\ref{fig:schematic} and Sec.~\ref{sec:qip}). 
As with regular DMFT, the local self-energy scheme in this ``impurity-
DMFT'' represents a rather sophisticated approximation (Sec.~\ref{sec:locSE}),
and its application here produces nontrivial results.
Even though the real-space two-impurity model and parallel quantum dot models  are among the simplest examples for which DMFT can be used, they 
are in fact already at the limit of complexity for an exact
treatment with NRG. These models therefore present
an almost unique opportunity to test the validity of the
local self-energy approximation by comparing exact NRG
and impurity-DMFT results. 

In the context of quantum impurity models, Ref.~\onlinecite{titvinidze2012dynamical,*titvinidze2013boundary}
focused on the two-impurity Anderson model in a one-dimensional geometry. Calculation of various static spin susceptibilities (and comparison
with DMRG results) showed that the impurity-DMFT is quantitatively
correct at large impurity separations. However, the impurity-DMFT
clearly breaks down at shorter distances, when it predicts a transition to a symmetry-broken state.\cite{titvinidze2012dynamical,*titvinidze2013boundary}

Here we focus on dynamical quantities; in particular spectral functions, which are the central experimental observables for these systems. In the two-impurity model, we consider explicitly a three-dimensional (3D) cubic lattice host. The exact NRG solution discussed in Sec.~\ref{sec:2iam_nrg} shows a range of correlated electron physics, with RKKY-dominated effects for small impurity separations, and standard Kondo in the dilute impurity limit. Impurity-DMFT is correctly able to determine the onset of the dilute limit,\cite{mitchell2015multiple} and certain low-energy properties; however when the Kondo effect is suppressed by non-local inter-impurity RKKY coupling, impurity-DMFT fails qualitatively (see Sec.~\ref{sec:2iam_dmft}). On the other hand, impurity-DMFT is shown to perform surprisingly well for models of parallel coupled quantum dots, capturing for example nontrivial underscreened Kondo physics through the self-consistent generation of a critical pseudogapped effective medium in the local problem --- see Sec.~\ref{sec:QDs}. Magnetization effects on application of a dot field are also very well captured (Sec.~\ref{sec:mag}), although the quantum phase transition between high- and low-spin dot states on tuning inter-dot coupling is not describable within impurity-DMFT (Sec.~\ref{sec:t}). We conclude in Sec.~\ref{sec:conc} with a wider discussion of the applicability of inhomogeneous DMFT.


\section{Quantum impurity problems and exact solution with NRG}
\label{sec:qip}

Quantum impurity systems, involving either one or several coupled impurities, are modeled in terms of generalized Kondo or Anderson Hamiltonians.\cite{hewson1997kondo} They can be decomposed as
$H=H_{\text{host}} + H_{\text{imp}} + H_{\text{hyb}}$, where
$H_{\text{host}}$ and $H_{\text{imp}}$ describe the isolated host and
impurities, respectively; and $H_{\text{hyb}}$ describes the coupling
between them. In the following, we focus on models of $N$ coupled Anderson impurities. Each impurity $\alpha$ is regarded as a single spinful correlated quantum level, 
\begin{equation}
\label{eq:Himp}
H_{\mathrm{imp}} = \sum_{\alpha=1}^{N} \Big [ \sum_{\sigma} \epsilon^{\phantom{\dagger}}_{\alpha}\cre{d}{\alpha\sigma}\ann{d}{\alpha\sigma} + U^{\phantom{\dagger}}_{\alpha}\cre{d}{\alpha\uparrow}\ann{d}{\alpha\uparrow}\cre{d}{\alpha\downarrow}\ann{d}{\alpha\downarrow}\Big ] + H_{\text{imp-imp}}\;, 
\end{equation}
where $\cre{d}{\alpha\sigma}$ creates an electron on level $\alpha=1, ... , N$ with spin $\sigma=\uparrow/\downarrow$. The single-particle level energy is $\epsilon_{\alpha}$, while $U_{\alpha}$ is the local Coulomb repulsion penalizing double occupancy. Additionally, there may be inter-impurity couplings, embodied by $H_{\text{imp-imp}}$. In the case of coupled quantum dots, 
\begin{equation}
\label{eq:imp-imp}
H_{\text{imp-imp}}=\sum_{\alpha\ne\beta,\sigma} v^{\phantom{\dagger}}_{\alpha\beta}\cre{d}{\alpha\sigma}\ann{d}{\beta\sigma} + \text{H.c} \;,
\end{equation}
where $v_{\alpha\beta}$ is a tunnel-coupling matrix element connecting different dots $\alpha$ and $\beta$. For real magnetic impurity clusters on surfaces, Eq.~\ref{eq:imp-imp} describes a direct intra-cluster coupling. In modeling multilevel quantum dots, the different levels are regarded as different `impurities'; then $H_{\text{imp-imp}}$ may include a ferromagnetic Hund's rule exchange between levels.\cite{logan2009correlated} Capacitive couplings between different dots could also be included.

\begin{figure}[t]
\begin{center}
\includegraphics[width=80mm]{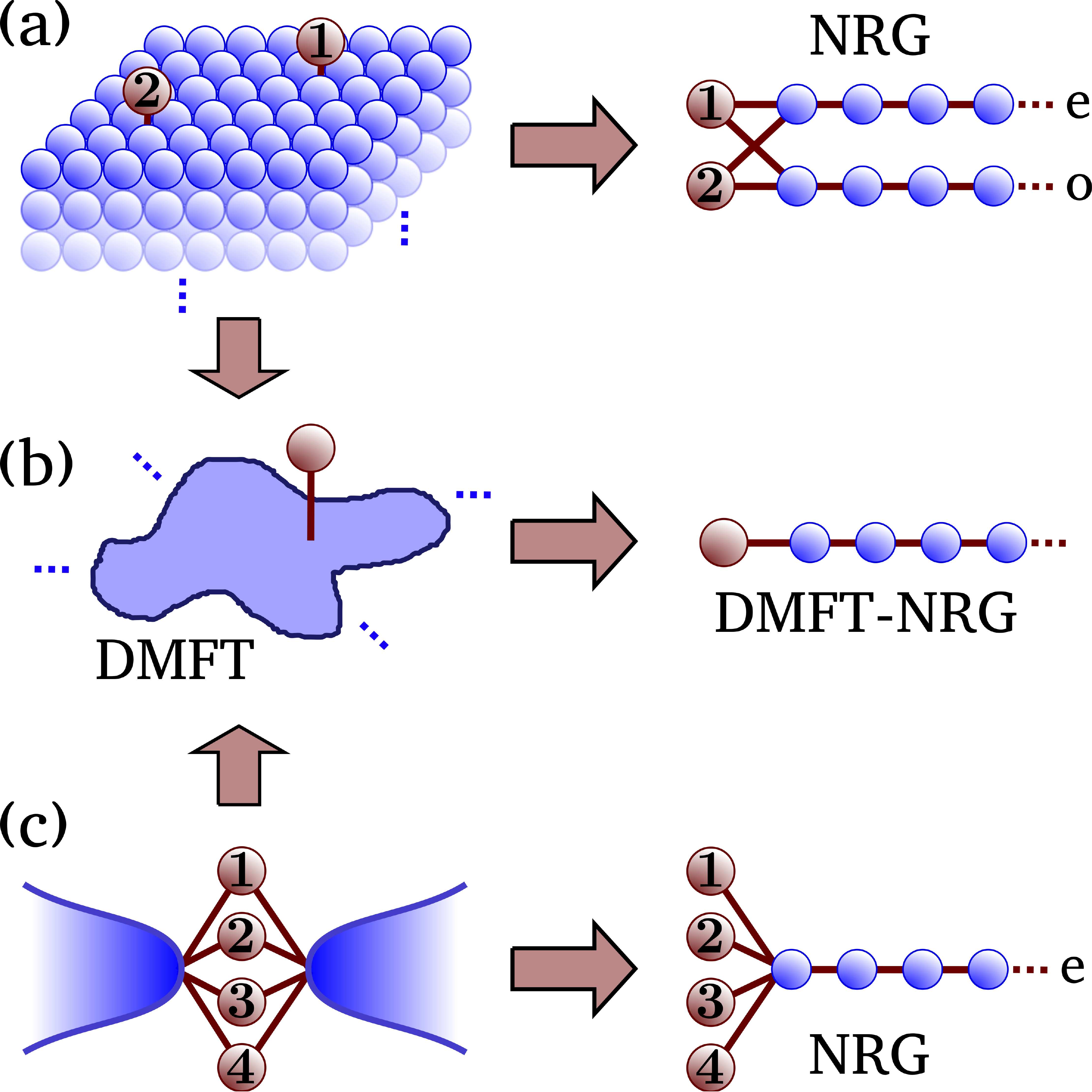}
\caption{\label{fig:schematic}
Schematic of the models and mappings used. (a) Two separated impurities on the surface of a 3D cubic lattice, mapped to a problem involving two Wilson chains in NRG. (b) Mapping to a single-impurity problem in a self-consistently determined effective medium, via the local self-energy approximation (``impurity-DMFT''). (c) Quantum dots coupled in parallel between source and drain leads, mapped to a multi-impurity but single-channel problem in NRG.}
\end{center}
\end{figure}

The host metal is taken to be noninteracting; in a diagonal basis it is described generically by,
\begin{equation}
\label{eq:Hhost_diag}
H_{\text{host}} = \sum_{k,\sigma} \epsilon^{\phantom{\dagger}}_k \cre{c}{k\sigma}\ann{c}{k \sigma} \;.
\end{equation}
In the thermodynamic limit, the conduction electrons form a continuous band of half-width $D$, and total density of states per spin $\rho_{\text{tot}}(\omega)=\sum_k \delta(\omega-\epsilon_k)$.

The hybridization between the impurities and conduction electrons is given by,
\begin{equation}
\label{eq:Hhyb}
H_{\text{hyb}} = \sum_{\alpha}\sum_{k,\sigma} V^{\phantom{\dagger}}_{\alpha k} \cre{d}{\alpha\sigma}\ann{c}{k \sigma} + \text{H.c.} \;.
\end{equation}

Note that the corresponding Kondo models are obtained via a Schrieffer-Wolff transformation\cite{hewson1997kondo} in the limit of single occupation for each level, perturbatively eliminating excitations involving empty or doubly-occupied levels to second order in $H_{\text{hyb}}$.

The question of how many electronic screening \emph{channels} are involved in such models is a subtle one. Although in general one can expect $N_c=N$ screening channels for a system of $N$ impurities, in fact any $1\le N_c \le N$ can be realized, depending on the parameters $V_{\alpha k}$. 

Consider the $N\times N$ impurity Green function matrix $\mathbf{G}_{\sigma}^{\text{imp}}(\omega)$, which contains all the information on the impurity single-particle dynamics. In general, there are finite local and non-local elements 
$G_{\alpha\beta,\sigma}^{\text{imp}} = \langle \langle \ann{d}{\alpha\sigma};\cre{d}{\beta\sigma}\rangle\rangle_{\omega}$  (as usual,
$\langle \langle \hat{A};\hat{B}\rangle\rangle_{\omega}$ is the
Fourier transform of the retarded correlator $-i\theta(t)\langle \{
\hat{A}(t),\hat{B}(0) \}_{+}\rangle$). The matrix Dyson equation then reads,  
\begin{equation}
\label{eq:dyson} 
\left[\mathbf{G}_{\sigma}^{\mathrm{imp}}(\omega)\right]^{-1}=\left[\mathbf{g}_{\sigma}^{\mathrm{imp}}(\omega)\right]^{-1}-\mathbf{\Sigma}_{\sigma}(\omega) \;.
\end{equation} 
Here, $\mathbf{\Sigma}_{\sigma}(\omega)$ is the interaction self-energy matrix, containing all of the nontrivial correlated electron physics of the problem. The noninteracting (but host-coupled) impurity Green functions are given by
\begin{equation} 
\label{eq:gimp}
\left[\mathbf{g}_{\sigma}^{\mathrm{imp}}(\omega)\right]^{-1}=(\omega+i0^{+})\mathbf{I}-\textbf{E}-\textbf{v}-\mathbf{\Gamma}(\omega) \; ,
\end{equation} 
where $[\textbf{E}]_{\alpha\beta}=\delta_{\alpha\beta}\epsilon_{\alpha}$, $[\textbf{v}]_{\alpha\beta}=v_{\alpha\beta}$, and elements of the hybridization matrix are given by  $\Gamma_{\alpha\beta}(\omega)=\sum_k V^{*}_{\alpha k}V^{\phantom{*}}_{\beta k} / (\omega+i0^{+}-\epsilon_k)$.

In certain high-symmetry cases, the hybridization matrix $\mathbf{\Gamma}(\omega)$ can be diagonalized by a single canonical transformation of operators (that is, \emph{independently} of energy $\omega$). A given channel in this basis is then decoupled from the impurity subsystem if it has zero eigenvalue, $\tilde{\Gamma}_{ii}(\omega)=0$, yielding $N_c<N$.

For $N_c=N$ one generally expects an overall singlet ground state, with each impurity eventually fully Kondo screened. By contrast, for $N_c<N$ there is the possibility of \emph{underscreening}, in which the ground state remains degenerate. We consider examples of each below; see also Fig.~\ref{fig:schematic}.


\subsection{Two-impurity problem in real-space}
\label{sec:2iam_model}

When two Anderson impurities $\alpha=1$ and $2$ are embedded in a host metal at different real-space positions $\textbf{r}_1$ and $\textbf{r}_2$, a rich range of correlated electron physics can result due to the competition between the Kondo effect and a through-host RKKY coupling.\cite{jones1987study,*jones1988low,sakai1992excitation,silva1996particle,allerdt2015kondo,mitchell2015multiple,schwabe2015screening} 

We will assume here that there is no \emph{direct} inter-impurity coupling, $H_{\text{imp-imp}}=0$. We take as a concrete physical example, a semi-infinite 3D cubic tight-binding host with a (100) surface,
\begin{equation}
\label{eq:Hhost}
H^{3D}_{\mathrm{host}} = -t\sum_{\langle ij\rangle,\sigma}(\cre{c}{\textbf{r}_i\sigma}\ann{c}{\textbf{r}_j\sigma} + \text{H.c.}) \;,
\end{equation}
where $t$ is the tunneling matrix element connecting nearest-neighbor sites 
$i$ and $j$. 
This simplified real-space realization captures many features of real
metals --- in particular, there is a definite bandwidth $2D=12t$, and the
surface density of states is finite and flat around the Fermi level, with $\rho_{\text{loc}}(\omega=0) = 1/(6t)$. Eq.~\ref{eq:Hhost_diag} is the diagonal representation of Eq.~\ref{eq:Hhost}.

In this real-space basis, the hybridization Hamiltonian, Eq.~\ref{eq:Hhyb}, reads,
\begin{equation}
\label{eq:AIM_hyb}
H^{\textit{2IAM}}_{\mathrm{hyb}} = \sum_{\alpha=1,2}\sum_{\sigma} V_{\alpha} \cre{d}{\alpha\sigma}\ann{c}{\textbf{r}_{\alpha}\sigma} + \text{H.c.}  \; . 
\end{equation}
Elements of the hybridization matrix then follow as 
$\Gamma_{\alpha\beta}(\omega) = V^{*}_{\alpha}V^{\phantom{*}}_{\beta} G^0(\textbf{r}_{\alpha},\textbf{r}_{\beta},\omega) $, where $G^0(\textbf{r}_{\alpha},\textbf{r}_{\beta},\omega)=\langle \langle \ann{c}{\textbf{r}_{\alpha}\sigma};\cre{c}{\textbf{r}_{\beta}\sigma}\rangle\rangle^0_{\omega}$ is the propagator between
sites $\textbf{r}_{\alpha}$ and $\textbf{r}_{\beta}$ of the clean host
(without impurities).  Due to translational invariance on the surface of the clean host, $\Gamma_{11}(\omega)=\Gamma_{22}(\omega)\equiv \Gamma_{\text{loc}}(\omega)$ and $\Gamma_{12}(\omega)=\Gamma_{21}(\omega)$. Note that the non-local functions $\Gamma_{\alpha\ne\beta}(\omega)$ depend on the inter-impurity separation $\textbf{R}=\textbf{r}_1-\textbf{r}_2$, and must be computed explicitly for each real-space realization of the two-impurity problem.
Here we compute the lattice Green functions numerically exactly using the  convolution method of Ref.~\onlinecite{derry2015quasiparticle}, which can be formulated for systems with a surface.

We take now \emph{equivalent} impurities, $\epsilon_{\alpha}\equiv \epsilon$, $U_{\alpha}\equiv U$, and $V_{\alpha}\equiv V$. 
By transforming to an even/odd impurity orbital basis $d_{e/o,\sigma}=\tfrac{1}{\sqrt{2}}(d_{1 \sigma}\pm d_{2 \sigma})$,
the $2\times 2$ matrix Dyson equation, Eq.~\ref{eq:dyson}, is diagonalized to yield,
\begin{equation}
\label{eq:dyson_e/o} 
\left [ G_{e/o,\sigma}^{\mathrm{imp}}(\omega) \right ]^{-1}=\omega+i0^{+}-\epsilon-\Gamma_{e/o}(\omega) - \Sigma_{e/o,\sigma}(\omega) \;, 
\end{equation} 
where $\Omega_{e/o}(\omega)=\Omega_{11}(\omega)\pm\Omega_{12}(\omega)$ for $\Omega=G$, $\Gamma$, and $\Sigma$. The even and odd channels are strictly decoupled in the non-interacting ($U=0$) system: the even(odd) impurity combination couples only to the even(odd) conduction electron channel. 
However, note that $\Sigma_{e}(\omega)$ and $\Sigma_{o}(\omega)$ each contains information on \emph{both} impurities, and their mutual correlations.

Since $\Gamma_e(\omega)\ne \Gamma_o(\omega) \ne 0$, the full real-space two-impurity Anderson model (2IAM) is irreducibly two-channel; see Fig.~\ref{fig:schematic}(a).


\subsection{Coupled quantum dots}
\label{sec:CQD_model}

Several coupled quantum dots, connected between source and drain metallic leads, are modeled by Eqs.~\ref{eq:Himp}--\ref{eq:Hhyb}. Here we focus on the specific geometry of quantum dots in \emph{parallel}\cite{vzitko2006multiple,*bonvca2007quantum} [see  Fig.~\ref{fig:schematic}(c)], which also mimics a multilevel dot.\cite{logan2009correlated} We consider explicitly the symmetric setup with $\epsilon_{\alpha}\equiv \epsilon$, $U_{\alpha}\equiv U$, and $v_{ij}\equiv v$. The dot-lead hybridization for a system of $N$ dots is taken to be,
\begin{equation}
\label{eq:Hhyb_CQD}
H_{\text{hyb}}^{\textit{PQD}} = V\sum_{\alpha=1}^{N}\sum_{\sigma} \cre{d}{\alpha\sigma}\ann{c}{0 \sigma} + \text{H.c.}  \; ,
\end{equation}
where $\ann{c}{0 \sigma}$ is an operator for the symmetric combination of lead orbitals coupling to the dots. The other conduction electron states couple indirectly to the dots via $\ann{c}{0 \sigma}$. 

The impurity Green functions for this parallel quantum dot (PQD) model are given by the $N\times N$ matrix Dyson equation, Eq.~\ref{eq:dyson}, with elements of the hybridization matrix in Eq.~\ref{eq:gimp} identical by symmetry, $\Gamma_{\alpha\beta}(\omega) \equiv \Gamma_{\text{loc}}(\omega)$. For simplicity, in the following we take $\Gamma_{\text{loc}}(\omega)$ to be the same as in Sec.~\ref{sec:2iam_model}.

In fact, only the symmetric (`even') combination of dot orbitals couples directly to the leads,
\begin{equation}
\label{eq:Hhyb_CQD_e}
H_{\text{hyb}}^{\textit{PQD}} = \sqrt{N}\times V\sum_{\sigma} \cre{d}{e\sigma}\ann{c}{0 \sigma} + \text{H.c.}  \; ,
\end{equation}
with $d_{e\sigma}=N^{-1/2}\sum_{\alpha=1}^N d_{\alpha\sigma}$. The other dot orbital combinations couple indirectly through $d_{e\sigma}$. 
The matrix Dyson equation can again be diagonalized, yielding in particular for the even-orbital Green function $G_{e,\sigma}^{\text{imp}} = \langle \langle \ann{d}{e\sigma};\cre{d}{e\sigma}\rangle\rangle_{\omega}$,
\begin{equation}
\label{eq:CQD_dyson_e} 
\left [ G_{e,\sigma}^{\mathrm{imp}}(\omega) \right ]^{-1}= \omega+i0^{+}-\epsilon-(N-1)v-\Gamma_{e}(\omega) - \Sigma_{e,\sigma}(\omega)   \;, 
\end{equation}
where $\Gamma_e(\omega)= N\Gamma_{\text{loc}}(\omega)$. The other $(N-1)$ conduction electron channels decouple from the impurities, $\Gamma_o(\omega)=\Gamma_{\alpha\alpha}(\omega)-\Gamma_{\alpha\ne\beta}(\omega)=0$, leaving a \emph{single-channel} description. Note, however, that the self-energy $\Sigma_e(\omega)=\Sigma_{\alpha\alpha}(\omega)+(N-1)\Sigma_{\alpha\ne\beta}(\omega)$ still inherently contains inter-impurity correlations.


\subsection{Numerical renormalization group}
\label{sec:nrg}

A wide range of quantum impurity problems can be solved exactly using NRG,\cite{bulla2008numerical} including the above real-space 2IAM and PQD models.

The NRG method involves mapping the full quantum impurity model to a 1D form amenable to iterative diagonalization. Specifically, the impurities must be coupled to the end of one or more \emph{decoupled} Wilson chains --- see Fig.~\ref{fig:schematic}. This necessitates a diagonal structure of the hybridization matrix $\Gamma_{\alpha\beta}(\omega)\propto \delta_{\alpha\beta}$, each nonzero element of which is discretized logarithmically, and mapped to a semi-infinite 1D tight-binding Wilson chain.\cite{bulla2008numerical}  
Starting from the impurity subsystem, Wilson orbitals are successively coupled on, and the Hamiltonian defined in the larger Hilbert space is diagonalized. Only the $N_K$ lowest energy states are retained for construction of the Hamiltonian at the next step. This truncation scheme is justified by the exponential decay of hoppings down the Wilson chain. Successively lower energies are reached at each step, revealing the RG character of the problem.

In the case of the real-space 2IAM, $\boldsymbol{\Gamma}(\omega)$ is diagonal in the even/odd orbital basis, with
$\Gamma_{e/o}(\omega)=\Gamma_{11}(\omega)\pm\Gamma_{12}(\omega)$, such that  information about the real-space separation of the impurities is encoded in the \emph{difference} between $\Gamma_{e}(\omega)$ and $\Gamma_{o}(\omega)$. Discretizing each and mapping to even and odd Wilson chains leads to the setup illustrated in Fig.~\ref{fig:schematic}(a), with two impurities and two channels. In our NRG implementation, we discretize $\Gamma_{e/o}(\omega)$ using $\Lambda=2.5$, label states by total charge and spin projection quantum numbers to block diagonalize the Hamiltonian, and retain $N_K=15000$ states at each step. So-called $z$-averaging was not required.

Generalized two-channel problems of this type are computationally demanding.\cite{mitchell2014generalized,iNRG} Furthermore, the local environments of the two impurities must be identical to accomplish the above mapping. If, for example, the impurities are different distances away from the surface, or if the host is disordered, then $\boldsymbol{\Gamma}(\omega)$ cannot simply be diagonalized by canonical transformations. Similarly, in the case of three impurities, NRG can only treat a symmetric geometry with $\Gamma_{\alpha\alpha}(\omega)=\Gamma_{\text{loc}}(\omega)$ and $\Gamma_{\alpha\ne\beta}(\omega)=\Gamma_{\text{non-loc}}(\omega)$. The resulting three-channel problem would also require significantly greater computational resources than that of the two-impurity case. 

In the case of PQD models, the formulation in Sec.~\ref{sec:CQD_model} is single-channel: $\Gamma_e(\omega)$ can be discretized and mapped to a Wilson chain with the dots at one end, as illustrated in Fig.~\ref{fig:schematic}(c). The complexity here is contained in the impurity Hamiltonian itself, which involves $N$ interacting orbitals. In practice this limits NRG to models with $N\lesssim 8$. The possibility of Kondo underscreening and frustration due to finite interdot coupling $v$ means that a rather large number of states $N_K=5000-8000$ must be kept at each step to accurately capture the low-energy physics (for $\Lambda=2$).

Impurity spectral functions presented in the following sections are obtained from the matrix Dyson equation, Eq.~\ref{eq:dyson}, using the appropriate exact non-interacting propagators $\textbf{g}^{\text{imp}}_{\sigma}(\omega)$. The nontrivial physics enters through the $N\times N$ self-energy matrix, which is calculated here using NRG via a generalization of Ref.~\onlinecite{bulla1998numerical}, from the full density matrix\cite{weichselbaum2007sum} constructed in the complete Anders-Schiller basis.\cite{anders2005real}


\section{Local self-energy approximation for quantum impurity problems}
\label{sec:locSE}

While the above models can still be solved exactly by standard NRG, they are computationally demanding, and represent an approximate limit on what can be achieved at present. Clearly true many-impurity systems and complex nanostructures are out of reach using full NRG. This provides motivation for development of approximate schemes capable of dealing with Kondo physics in multi-impurity systems.

An inhomogeneous (or real-space) DMFT approach has been widely used to treat generalized Hubbard models which lack translational invariance.\cite{potthoff1999surface,*potthoff1999metallic} At its heart, it involves a local self-energy approximation, which allows the full problem to be mapped onto a set of auxiliary single-impurity Anderson models (one for each inequivalent site in the system), each in an effective medium that must be determined self-consistently. Quantum impurity problems can in fact be regarded as highly inhomogeneous Hubbard models, and we apply a similar local self-energy approximation (or ``impurity-DMFT'') here to the real-space 2IAM and PQD models --- see Fig.~\ref{fig:schematic}.

We first recap briefly the generalized method. The impurity Green functions are obtained from the matrix Dyson equation Eq.~\ref{eq:dyson} using the exact non-interacting propagators $\textbf{g}_{\sigma}^{\text{imp}}(\omega)$. However, within impurity-DMFT the self-energy matrix takes the form,
\begin{equation}
\label{eq:local_SE}
\left [ \mathbf{\Sigma}_{\sigma}(\omega) \right ]_{ij} ~\overset{\text{DMFT}}{=}~ \delta_{ij}\Sigma^{\text{loc}}_{i \sigma}(\omega) \;.
\end{equation}

For each inequivalent site $\gamma$, we introduce an auxiliary Anderson impurity model,
\begin{equation}
\label{eq:AIM}
\begin{split}
H^{\gamma}_{\textit{aux}} = &\sum_{\sigma}\epsilon^{\phantom{\dagger}}_{\gamma}\cre{d}{\sigma}\ann{d}{\sigma} + U^{\phantom{\dagger}}_{\gamma}\cre{d}{\uparrow}\ann{d}{\uparrow}\cre{d}{\downarrow}\ann{d}{\downarrow}\\
&+ \sum_{k,\sigma}\tilde{\epsilon}^{\phantom{\dagger}}_{\gamma k} \cre{c}{k\sigma}\ann{c}{k\sigma}+\sum_{k,\sigma} (\tilde{V}^{\phantom{\dagger}}_{\gamma k} \cre{d}{\sigma}\ann{c}{k\sigma}+ \text{H.c.} ) \;,
\end{split}
\end{equation}
where the parameters $\epsilon_{\gamma}$ and $U_{\gamma}$ are specified by the original quantum impurity model. $\tilde{\epsilon}_{\gamma k}$ and $\tilde{V}_{\gamma k}$ are however determined self-consistently. All the relevant information about the effective medium to which the impurity couples is contained in the hybridization function $\Delta_{\gamma}(\omega)=\sum_k |\tilde{V}_{\gamma k}|^2/(\omega+i0^{+}-\tilde{\epsilon}_{\gamma k})$. The impurity Green function of this auxiliary model is therefore given by,
\begin{equation}
\label{eq:AIM_G}
[\mathcal{G}^{\textit{aux}}_{\gamma \sigma}(\omega)]^{-1} = \omega+i0^+ -\epsilon_{\gamma} -\Delta_{\gamma}(\omega)- \Sigma_{\gamma \sigma}^{\textit{aux}}(\omega) \;.
\end{equation}
The self-energy $\Sigma_{\gamma \sigma}^{\textit{aux}}(\omega)$ is then calculated (e.g. using standard single-site single-channel NRG) for each inequivalent site $\gamma$, given the hybridization function $\Delta_{\gamma}(\omega)$. 

$\Delta_{\gamma}(\omega)$ is determined self-consistently on the level of the local Green functions by demanding:
\begin{subequations}
\label{eq:SC}
\begin{align}
G_{\gamma\gamma,\sigma}^{\text{imp}}(\omega) &= \mathcal{G}^{\textit{aux}}_{\gamma \sigma}(\omega) \;, \\
\Sigma^{\text{loc}}_{\gamma\sigma}(\omega) &= \Sigma_{\gamma \sigma}^{\textit{aux}}(\omega) \;.
\end{align}
\end{subequations}
The self-consistency loop is as follows:\\
(1) Guess an initial form for $\Delta_{\gamma}(\omega)$.\\
(2) Calculate $\Sigma_{\gamma\sigma}^{\textit{aux}}(\omega)$ for auxiliary model Eq.~\ref{eq:AIM}, using $\Delta_{\gamma}(\omega)$ as input. Repeat for all inequivalent sites $\gamma$.\\
(3) Equate $\Sigma^{\text{loc}}_{\gamma\sigma}(\omega) = \Sigma_{\gamma\sigma}^{\textit{aux}}(\omega)$, construct $\boldsymbol{\Sigma}_{\sigma}(\omega)$ from Eq.~\ref{eq:local_SE}, and then find $G_{\gamma\gamma,\sigma}^{\text{imp}}(\omega)$ from Eq.~\ref{eq:dyson}.\\
(4) Equate $G_{\gamma\gamma,\sigma}^{\text{imp}}(\omega) = \mathcal{G}_{\gamma\sigma}^{\textit{aux}}(\omega)$ and find a new $\Delta_{\gamma}(\omega)$ via Eq.~\ref{eq:AIM_G}.\\
(5) Repeat steps (2)--(4) until converged (i.e. $G_{\gamma\gamma,\sigma}^{\text{imp}}(\omega)$ does not change upon further iteration).\\

An important feature of this self-consistency, which is perhaps not immediately obvious, is that it depends crucially on the \emph{basis} used for Eq.~\ref{eq:dyson}. 

For example, in the real-space 2IAM, different solutions can be converged, depending on whether one uses $G_{ij,\sigma}^{\text{imp}}(\omega)$ with $i$ and $j$ labeling either the physical impurity sites $1$ and $2$, or their even and odd orbital combinations. Indeed, in the latter case, there is no self-consistency loop at all: comparing Eqs.~\ref{eq:dyson_e/o} and \ref{eq:AIM_G} yields directly $\Delta_{e/o}(\omega)=\Gamma_{e/o}(\omega)$ for the inequivalent even and odd impurities. The local Green functions for the physical impurities (which are identical by symmetry) follow simply as the average,  $G_{11,\sigma}^{\text{imp}}(\omega)=\tfrac{1}{2}[G_{e,\sigma}^{\text{imp}}(\omega)+G_{o,\sigma}^{\text{imp}}(\omega)]$.

By contrast, establishing self-consistency directly on the level of $G_{11,\sigma}^{\text{imp}}(\omega)$ implies that $\Delta_{1}(\omega)=\Gamma_{11}(\omega)+[\Gamma_{12}(\omega)]^2/[\omega+i0^{+}-\epsilon-\Gamma_{11}(\omega)-\Sigma_{1\sigma}^{\text{loc}}(\omega)]$. Iterating the DMFT cycle then produces nontrivial structure in $\Delta_{1}(\omega)$ due to feedback from the self-energy. This is the formulation employed in Sec.~\ref{sec:2imp}.

Similarly, for the PQD models studied using impurity-DMFT in Sec.~\ref{sec:QDs}, we establish self-consistency for the local Green functions in the physical impurity basis. 

In this work, we use NRG\cite{bulla2008numerical} to solve the auxiliary Anderson model, Eq.~\ref{eq:AIM}. For all impurity-DMFT calculations presented, we use $\Lambda=2$ and $N_K=3000$.


\section{Real-space two-impurity model}
\label{sec:2imp}

We examine now the real-space 2IAM, with two impurities located at different sites on the (100) surface of a 3D cubic lattice [Eqs.~\ref{eq:Himp}, \ref{eq:Hhost} and \ref{eq:AIM_hyb} with $N=2$; see Fig.~\ref{fig:schematic}(a)]. For simplicity we focus on the particle-hole symmetric case, $\epsilon=-U/2$. We recap first the true physics of the model, with exact numerical results obtained from the full NRG solution, before comparing with impurity-DMFT results. Fig.~\ref{fig:comp_2imprs} shows a comparison for the impurity spectral functions, with exact results as solid black lines, and impurity-DMFT as red dashed lines.


\subsection{Full NRG solution}
\label{sec:2iam_nrg}

\begin{figure}[t]
\begin{center}
\includegraphics[width=85mm]{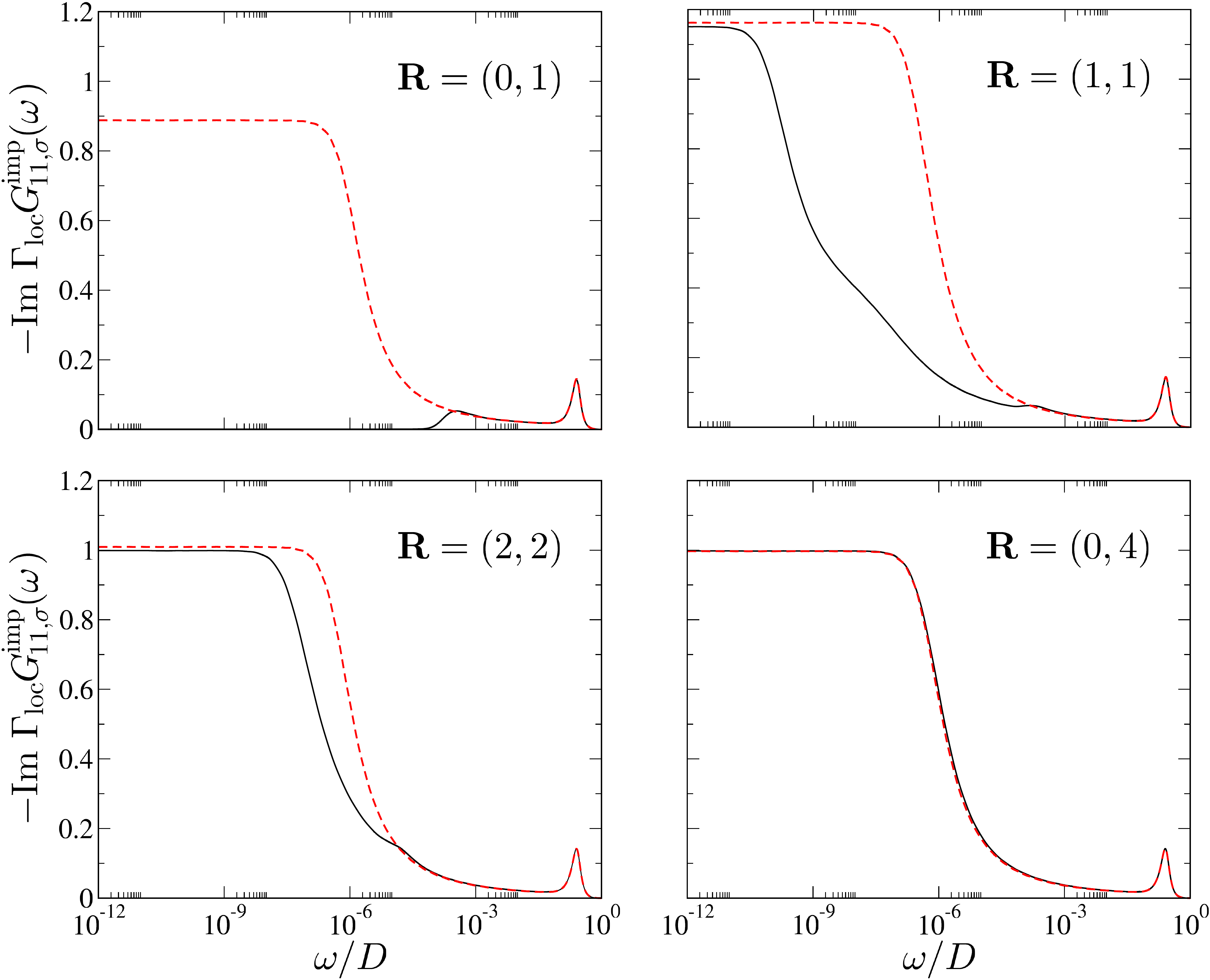}
\caption{\label{fig:comp_2imprs}
Comparison between exact NRG results (solid black lines) and impurity-DMFT results (dashed red lines) for the real-space 2IAM. The local impurity spectral function at $T=0$ is plotted as $-\text{Im}~\Gamma_{\text{loc}} G_{11,\sigma}^{\text{imp}}(\omega)$ vs $\omega/D$. The two impurities are taken to be on the (100) surface of a (semi-) infinite 3D cubic lattice, separated by the vector $\textbf{R}\equiv (R_x,R_y)$. Clockwise from top-left panel: $\textbf{R} =(0,1), (1,1), (0,4), (2,2)$. Impurity parameters: $U/D=0.5$, $\epsilon=-U/2$ and $V/D=0.075$.}
\end{center}
\end{figure}

At very large inter-impurity separations $|\textbf{R}|\rightarrow \infty$, each of the two impurities behaves independently and one recovers the well-known Kondo physics of the single-impurity Anderson model.\cite{hewson1997kondo} Indeed, the `dilute limit' of essentially independent Kondo screening sets in rather rapidly with increasing $|\textbf{R}|$ in 3D systems, as known from experiment\cite{CoonCu_Nikolaus,*CoonCu_Manoharan,*wahl_dens} and theory.\cite{mitchell2015multiple} 
In the dilute limit, the impurities are each screened by surrounding conduction electrons below an effective single-impurity Kondo temperature $T_K^{1\text{imp}}$. At low energies and temperatures, the impurity spectral resonance (experimentally observable in scanning tunneling spectroscopy\cite{CoonCu_Nikolaus,*CoonCu_Manoharan,*wahl_dens}) takes a universal Kondo form.\cite{hewson1997kondo}

For two impurities on the (100) surface of the 3D cubic lattice, the dilute limit is effectively realized for an inter-impurity separation $\textbf{R}\equiv (R_x,R_y)=(0,4)$ (with lattice constant $a_0\equiv 1$). The bottom-right panel of Fig.~\ref{fig:comp_2imprs} shows the corresponding impurity spectral function, which is of single-impurity form. Note in particular that the Friedel sum rule\cite{hewson1997kondo} (FSR) implies spectral pinning in the \emph{single} impurity case $-\text{Im}~\Gamma_{\text{loc}} G_{\sigma}^{1\text{imp}}(\omega=0)=1$, which as such provides a measure of the independence of Kondo screening in problems with several impurities. Here, $\Gamma_{\text{loc}} = -\text{Im}~\Gamma_{\text{loc}}(0)$. For $\textbf{R}=(0,4)$, the FSR is very well satisfied by NRG results for the full two-impurity problem. Furthermore, the non-local propagator $G_{12,\sigma}^{\text{imp}}(\omega)$ is negligible for this separation (not shown), indicating vanishing inter-impurity correlations.

On bringing the impurities closer, however, genuine two-impurity effects manifest due to through-host RKKY correlations.\cite{silva1996particle} This has a marked effect on the impurity spectral functions, as seen clearly in the other panels of Fig.~\ref{fig:comp_2imprs}. In addition to differences in the detailed form of the impurity spectrum, note that its Fermi-level value depends on $\textbf{R}$, and is not pinned by the FSR. Indeed, the nontrivial behavior of $-\text{Im}~\Gamma_{\text{loc}} G_{11,\sigma}^{\text{imp}}(0)$ is well-known\cite{sakai1992excitation} from simplified models of the 2IAM; its origin is two-impurity contributions to the self-energy $\boldsymbol{\Sigma}_{\sigma}(\omega)$.

For separations $\textbf{R}=(1,1)$ and $(2,2)$, the effective RKKY coupling is ferromagnetic; the impurities bind together into a triplet state on the scale of $|J_{\text{RKKY}}|$, which is ultimately exactly Kondo-screened by the even and odd conduction electron channels. The RKKY scale can be read off from Fig.~\ref{fig:comp_2imprs} as the energy scale where the spectrum departs from single-impurity behavior. As expected, $J_{\text{RKKY}}$ is smaller in magnitude for $\textbf{R}=(2,2)$ than $(1,1)$. The anisotropy between even and odd channels in the case of $\textbf{R}=(1,1)$ is rather significant (due to large $\Gamma_{12}(\omega)$), leading to effective two-stage quenching of the inter-impurity triplet. For $\textbf{R}=(2,2)$, the even/odd anisotropy is weaker, and two-channel Kondo screening occurs essentially in a single step.

For neighboring impurities $\textbf{R}=(0,1)$, the effective RKKY interaction is antiferromagnetic. The impurities bind together into a \emph{singlet} state on the scale of $|J_{\text{RKKY}}|$, and decouple from the host, giving $G_{11,\sigma}^{\text{imp}}(0)\simeq 0$.

Note that in all cases, the overall ground state is a spin-singlet (there is no phase transition on varying parameters or impurity separation\cite{silva1996particle}). The lowest-energy physics is Fermi-liquid like, with 
$-\text{Im}\Sigma_{\alpha\beta,\sigma}(\omega)\sim \omega^2$ as $\omega\rightarrow 0$.


\subsection{Impurity-DMFT results}
\label{sec:2iam_dmft}

\begin{figure}[t]
\begin{center}
\includegraphics[width=70mm]{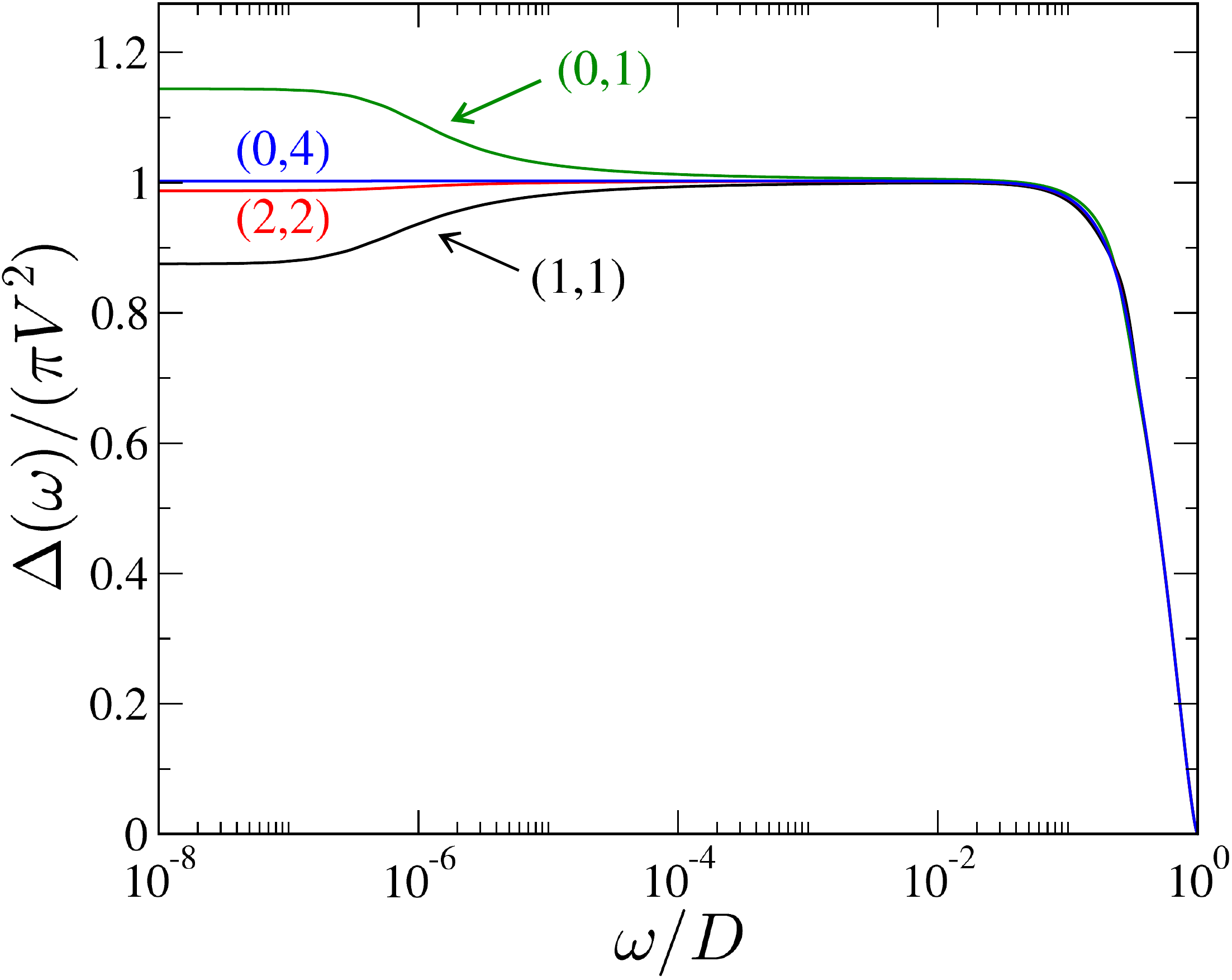}
\caption{\label{fig:hybrs}
Impurity-DMFT effective medium, $\Delta(\omega)/(\pi V^2)$ vs $\omega/D$ for the systems plotted in Fig.~\ref{fig:comp_2imprs}.
}
\end{center}
\end{figure}

We turn now to our impurity-DMFT results for the same systems --- see red dashed lines in Fig.~\ref{fig:comp_2imprs}. 

In the dilute limit, one expects effective single-impurity physics. The full self-energy matrix $\boldsymbol{\Sigma}_{\sigma}(\omega)$ then becomes diagonal, meaning purely \emph{local}. The impurity-DMFT scheme is exact in this limit. This is in fact seen in the bottom-right panel of Fig.~\ref{fig:comp_2imprs}, for impurity separation $\textbf{R}=(0,4)$: the exact NRG and impurity-DMFT solutions are perfectly in agreement. This is confirmed in Fig.~\ref{fig:hybrs}, where the DMFT effective medium $\Delta(\omega)$ is plotted. For $\textbf{R}=(0,4)$, we find under DMFT self-consistency that $\Delta(\omega)\simeq \Gamma_{\text{loc}}(\omega)$, meaning that the presence of the second impurity does not affect the local electronic environment of the first. Although this might seem trivial, it should be noted that the impurity-DMFT is correctly able to identify the onset of the dilute limit at \emph{finite} (and surprisingly small) impurity separations. Recovering single-impurity physics in impurity-DMFT is only strictly trivial as $|\textbf{R}|\rightarrow \infty$.

The impurity-DMFT scheme is however seen to break down as the impurities are brought closer together, and inter-impurity effects (embodied by non-local self-energy contributions) become important. The failure becomes increasingly severe on decreasing the separation.

For $\textbf{R}=(2,2)$ and $(1,1)$, the spectral behavior for energies $|\omega|\gg |J_{\text{RKKY}}|$ is correctly described within impurity-DMFT; but the effective two-channel screening of the inter-impurity triplet is not captured at lower energies. In particular, the two-stage screening in the case of $\textbf{R}=(1,1)$ is not apparent in the impurity-DMFT results, with the effect that the Kondo temperature (characterizing RG flow to the Fermi liquid fixed point) is significantly overestimated. $T_K$ is modified from the pure single-impurity result, but not sufficiently to correctly simulate the exact two-impurity results. 

However, it should be noted that in all Kondo-screened cases, the impurity-DMFT correctly recovers the nontrivial Fermi level value of the spectrum. This is achieved under DMFT self-consistency by the generation of a nontrivial effective medium, with $\Delta(\omega=0)$ deviating from $\Gamma_{\text{loc}}(\omega=0)$, as demonstrated in Fig.~\ref{fig:hybrs}.  This is consistent with Ref.~\onlinecite{titvinidze2012dynamical}, where an impurity-DMFT scheme was used to study two impurities in a 1D chain. In that work, static quantities at $T=0$ were considered, and shown to reproduce DMRG results for large separations. In 3D, as considered here, the dilute limit sets in much more rapidly. 
These results suggest that in the Kondo regime, quantities that depend only on $G_{11,\sigma}^{\text{imp}}(0)$, such as the $T=0$ resistivity due to impurities, might still be accessible with impurity-DMFT.

Finally, for neighboring impurities $\textbf{R}=(0,1)$, we see a catastrophic failure of impurity-DMFT. Here, non-local self-energy contributions are responsible for inter-impurity singlet formation, and cannot be captured by impurity-DMFT. In fact, impurity-DMFT predicts a Kondo-like resonance, with enhanced (rather than suppressed) spectral intensity at the Fermi energy.


\section{Coupled quantum dots}
\label{sec:QDs}

\begin{figure*}[t]
\begin{center}
\includegraphics[width=170mm]{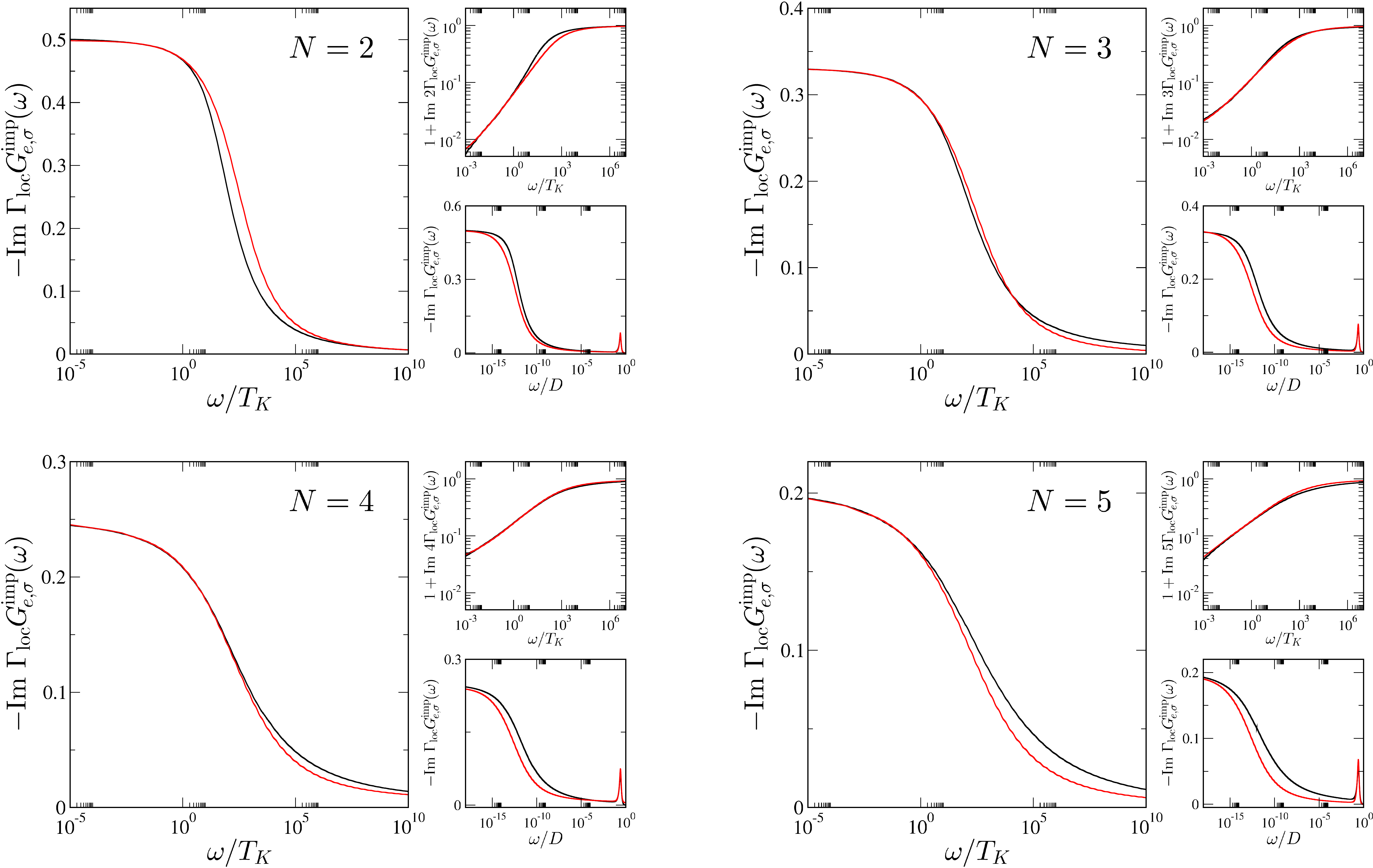}
\caption{\label{fig:comp_MLQD}
Comparison of impurity-DMFT results (red lines) and exact NRG (black lines) for PQD models with $N=2,3,4,5$. Main panel in each case shows the rescaled even orbital spectrum $-\text{Im}~\Gamma_{\text{loc}} G_{e,\sigma}^{\text{imp}}(\omega)$ vs $\omega/T_K$ at $T=0$, while the lower insets show the unscaled (raw) spectra. Upper insets show the detailed low-energy behavior. Here $T_K$ is defined via $\text{Im}~G_{e,\sigma}^{\text{imp}}(\omega=T_K)=0.95\times \text{Im}~G_{e,\sigma}^{\text{imp}}(\omega=0)$. Dot parameters: $U/D=0.5$, $\epsilon=-U/2$, $V/D=0.05$ (and $v=h=0$). 
}
\end{center}
\end{figure*}

We now discuss systems of $N$ quantum dots, coupled in parallel between two metallic leads, Eqs.~\ref{eq:Himp}-\ref{eq:Hhost_diag} and \ref{eq:Hhyb_CQD} --- see Fig.~\ref{fig:schematic}(c). Exact NRG results are surveyed in Fig.~\ref{fig:MLQD_ex_N}, while a detailed comparison between exact NRG and impurity-DMFT results is presented in Fig.~\ref{fig:comp_MLQD}.
In the following we use a comparatively large $U/D=0.5$ to allow for a detailed analysis of the universal low-energy physics. Although this parameter regime would be appropriate to describe the Kondo physics of e.g.\ 3d transition-metal impurities in break junctions, real quantum dot devices are typically characterized by a smaller $U/D$. We note that impurity-DMFT can also be applied to the latter case with equivalent accuracy.


\subsection{Full NRG solution}
\label{sec:PQDnrg}

\begin{figure}[t]
\begin{center}
\includegraphics[width=80mm]{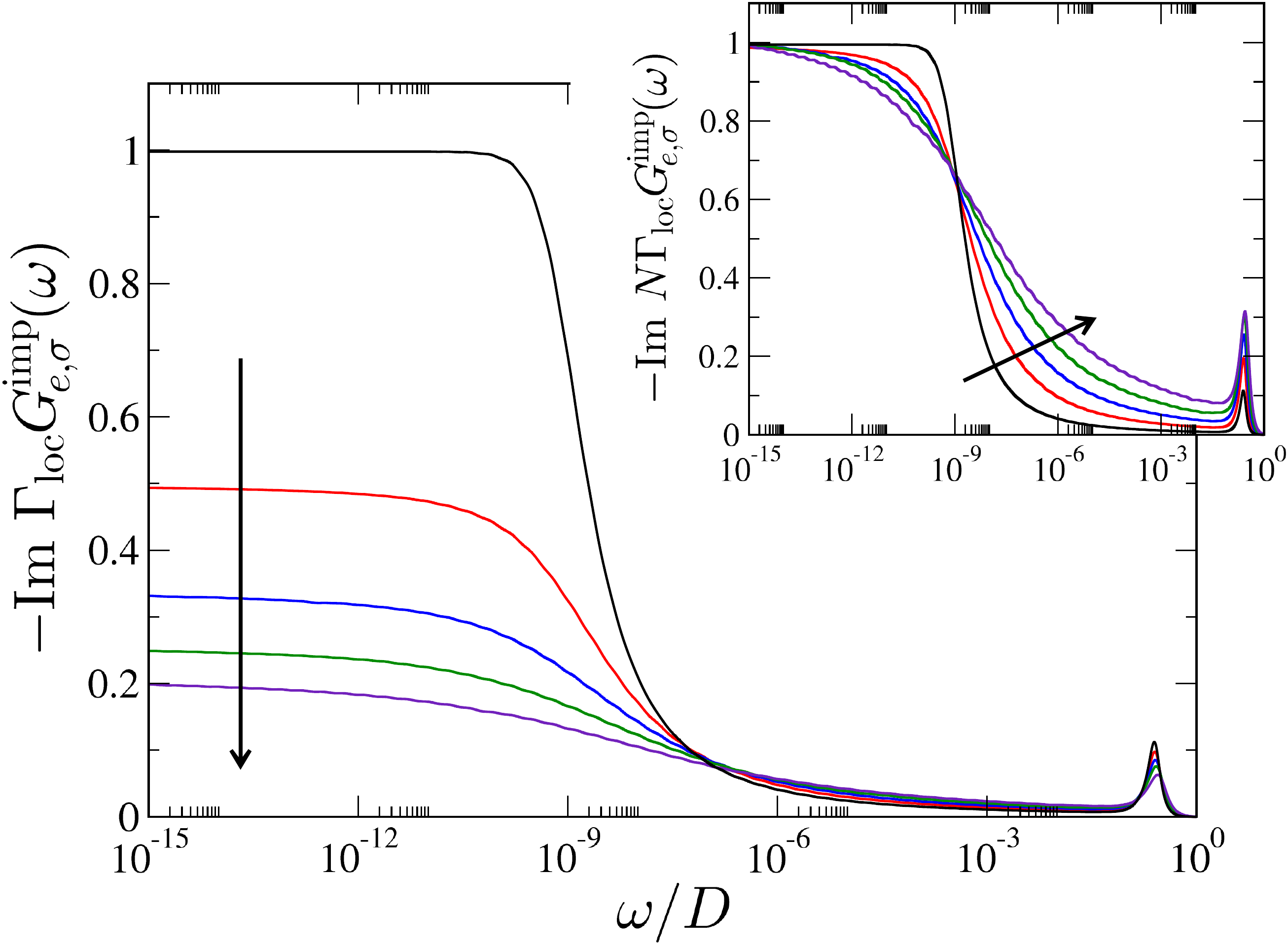}
\caption{\label{fig:MLQD_ex_N}
Exact NRG results for the PQD model consisting of $N$ coupled quantum dots. Plotted is the (even orbital) impurity spectral function $-\text{Im}~\Gamma_{\text{loc}} G_{e,\sigma}^{\text{imp}}(\omega)$ vs $\omega/D$ at $T=0$ for $N=1,2,3,4,5$ (black, red, blue, green and purple lines, following the arrow). Inset: rescaled spectra satisfying pinning condition, Eq.~\ref{eq:pin}. Impurity parameters: $U/D=0.5$, $\epsilon=-U/2$, $V/D=0.06$ (and $v=h=0$).}
\end{center}
\end{figure}

For a single quantum dot between source and drain leads ($N=1$), the standard Kondo effect is known to arise,\cite{cronenwett1998tunable,pustilnik2004kondo} producing a dot spectral function that takes a universal form when rescaled in terms of $T/T_K$ and $\omega/T_K$. The dot spin-$\tfrac{1}{2}$ degree of freedom in this case is \emph{exactly screened} by the lead conduction electrons below $T_K$, forming a ground state spin-singlet. The low-energy physics is described in terms of Fermi liquid theory,\cite{hewson1997kondo} which (in the case of particle-hole symmetry) implies,
\begin{subequations}
\begin{align}
\label{eq:FL}
\nonumber & \underline{N=1 :} \\
&\text{Im}~\Sigma_{11,\sigma}(\omega) \overset{\omega \ll T_K}{\sim} -(\omega/T_K)^2 \;,\\
\label{eq:FL_quad} &\text{Im}~\Gamma_{\text{loc}} \left [ G_{11,\sigma}^{\text{imp}}(\omega)-G_{11,\sigma}^{\text{imp}}(0) \right ]  \overset{\omega \ll T_K}{\sim} (\omega/T_K)^2 \;,\\
\label{eq:FL_FSR} &-\text{Im}~\Gamma_{\text{loc}} G_{11,\sigma}^{\text{imp}}(0) = 1 \;.
\end{align}
\end{subequations}
The FSR is embodied by Eq.~\ref{eq:FL_FSR}. This, and the characteristic quadratic approach of the spectrum to its Fermi level value Eq.~\ref{eq:FL_quad}, can be seen from exact NRG results in Fig.~\ref{fig:MLQD_ex_N} as the black line.

However, the physics of PQDs with $N>1$ (and similarly, multilevel dots\cite{sasaki2000kondo,logan2009correlated}) is rather different.\cite{vzitko2006multiple,*bonvca2007quantum,lu2005tunable,*de2006electronic}  In particular, an $\mathcal{O}(V^4)$ effective RKKY interaction arises between the dots, mediated by the conduction electrons,\cite{vzitko2006multiple,*bonvca2007quantum}
\begin{equation}
\label{eq:MLQD_JRKKY}
J_{\text{RKKY}} \sim - \Gamma_{\text{loc}}^2/U \;. 
\end{equation}
Importantly, this coupling is \emph{ferromagnetic}: on temperature/energy scales $\sim |J_{\text{RKKY}}|$, the $N$ dots form a combined spin-$N/2$ object. This spin-$N/2$ is coupled to a single channel of conduction electrons, and so is \emph{underscreened} to a spin-$(N-1)/2$ object by the Kondo effect below $T_K\ll J_{\text{RKKY}}$. The ground state therefore remains spin-degenerate\cite{mravlje2012ground} for $N>1$. The conduction electrons still feel a $\pi/2$ phase shift (at $T=0$ and $\omega=0$) due to the Kondo effect, resulting in a generalized pinning condition for the even orbital spectral function,
\begin{equation}
\label{eq:pin}
-\text{Im}~N\Gamma_{\text{loc}} G_{e,\sigma}^{\text{imp}}(0) = 1 \;.
\end{equation}
This is demonstrated in Fig.~\ref{fig:MLQD_ex_N}; the inset shows that Eq.~\ref{eq:pin} is satisfied numerically at low energies for all $N$. 

However, the full scaling form of the dot spectrum does depend on $N$ --- see Fig.~\ref{fig:MLQD_ex_N}. In particular, there are singular corrections\cite{koller2005singular} at low energies due to residual ferromagnetic correlations between the underscreened dot spin-$(N-1)/2$ and conduction electrons for $|\omega|\ll T_K$. This results in a `marginal Fermi liquid' at low energies, 
\begin{subequations}
\begin{align}
\label{eq:MFL}
\nonumber &\underline{N>1 :}\\
&\text{Im}~\Sigma_{e,\sigma}(\omega)\overset{\omega \ll T_K}{\sim} -\ln^{-2}(\omega/T_K) \;,\\
&\text{Im}~\Gamma_{\text{loc}} \left [ G_{e,\sigma}^{\text{imp}}(\omega)-G_{e,\sigma}^{\text{imp}}(0) \right ]  \overset{\omega \ll T_K}{\sim} \ln^{-2}(\omega/T_K) \;.
\end{align}
\end{subequations}
The low-energy spectral behavior is thus rather different for $N=1$ compared with $N>1$, reflecting the difference between exactly screened and underscreened Kondo.


\subsection{Impurity-DMFT results}

In the case of a single dot $N=1$, impurity-DMFT is exact by construction. In this subsection, we explore the extent to which the local self-energy approximation in impurity-DMFT can capture nontrivial underscreened Kondo physics in PQDs for $N>1$. Fig.~\ref{fig:comp_MLQD} compares exact NRG results (black lines) with impurity-DMFT results (red) for the even orbital spectral function in PQD models with $N=2,3,4,5$.

\begin{figure}[t]
\begin{center}
\includegraphics[width=60mm]{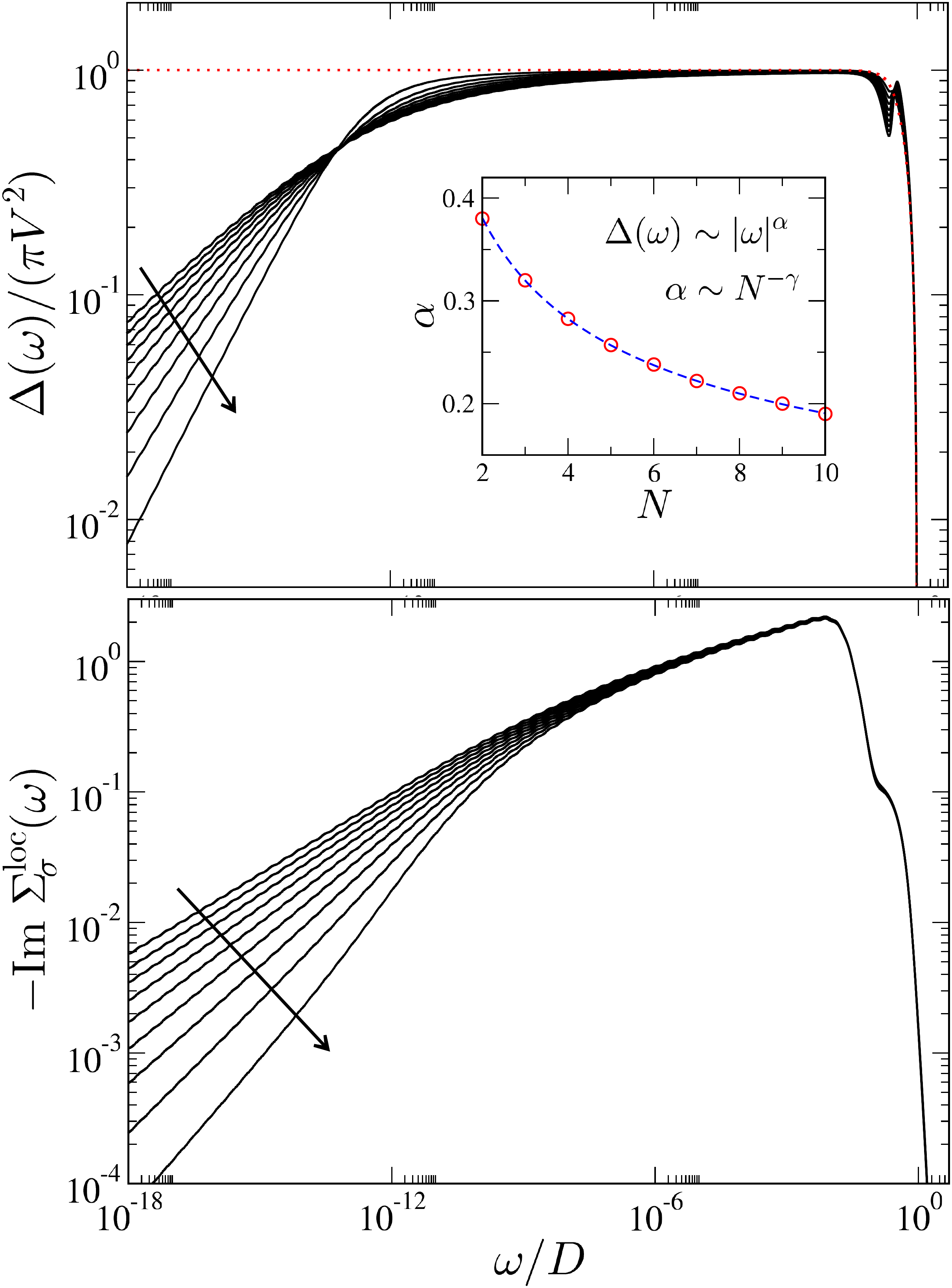}
\caption{\label{fig:hys_se}
Top panel: DMFT effective medium $\Delta(\omega)/(\pi V^2)$ vs $\omega/D$ for PQDs with $N=2,3,...,10$ increasing in the direction of the arrow. Low-energy behavior follows a pseudogap powerlaw $\Delta(\omega)\sim |\omega|^{\alpha}$, with exponent $\alpha\sim N^{-\gamma}$ that depends on the number of dots (see inset, points). We find numerically $\gamma\simeq 0.43$ (inset, dashed line). The bare hybridization is shown as the red dotted line for comparison.  Bottom panel: corresponding local DMFT self-energy $-\text{Im}~\Sigma_{\sigma}^{\text{loc}}(\omega)$, showing the same low-energy powerlaw decay as $\Delta(\omega)$. Same parameters as Fig.~\ref{fig:comp_MLQD}. 
}
\end{center}
\end{figure}

The impurity-DMFT solution is certainly seen to be approximate: the value of $T_K$ is always underestimated (lower insets of Fig.~\ref{fig:comp_MLQD}), and the precise scaling form is not entirely recovered (main panels). However, impurity-DMFT is apparently rather successful at capturing certain aspects of PQD physics. 

At higher energies $|\omega|\gg T_K$, one observes $1/\ln^2(\omega/T_K)$ spectral tails as expected. Furthermore, for $|\omega|\gg J_{\text{RKKY}}$, this behavior should be characteristic of single-impurity scattering. Perhaps it is not surprising the impurity-DMFT captures this physics asymptotically, since the PQD is mapped to an effective single-impurity model.

Far more interesting is the low-energy physics, which impurity-DMFT emulates surprisingly well in all cases considered (see upper insets of Fig.~\ref{fig:comp_MLQD} for each $N$). This is a rather nontrivial fact, since the exact results display marginal Fermi liquid physics\cite{koller2005singular} whose physical origin\cite{vzitko2006multiple,*bonvca2007quantum} is the Kondo underscreening of a high-spin object. This type of Kondo effect is obviously \emph{not} accessible within a single impurity Anderson model. 

Instead, the low-energy physics of the PQD model is mimicked within impurity-DMFT through generation of an effective medium with nontrivial low-energy structure. This is shown in the upper panel of Fig.~\ref{fig:hys_se} for different numbers of coupled dots, $N=2,3,...,10$ (increasing in the direction of the arrow). Although at high energies $|\omega|\gg T_K$, the DMFT effective medium $\Delta(\omega)\sim \Gamma_{\text{loc}}(\omega)$ is essentially that of the bare hybridization (dotted line), at lower energies $|\omega|\lesssim T_K$ a pseudogap develops in $\Delta(\omega)$. Specifically, we find the low-energy behavior to be well-described by,
\begin{equation}
\label{eq:pseudogap}
\Delta(\omega) ~\overset{|\omega|\ll T_K}{\sim}~ |\omega|^{\alpha} \;,
\end{equation}
where the exponent $\alpha\equiv \alpha(N)>0$ depends on the number of dots. Although there may be logarithmic corrections to this pure powerlaw behavior, we could not establish that from the numerical results. As shown in the inset to the upper panel of Fig.~\ref{fig:hys_se}, the exponent itself is found to follow a powerlaw form, $\alpha \sim N^{-\gamma}$ with $\gamma \simeq 0.43$. In particular, we note that $0<\alpha<\tfrac{1}{2}$ for all $N>1$. 

It is well-known that the particle-hole symmetric pseudogap Anderson model with $0<\alpha<\tfrac{1}{2}$ supports a quantum phase transition separating Kondo strong coupling and local moment phases.\cite{gonzalez1998renormalization,fritz2004phase,logan2014common} The critical point itself is interacting and is characterized by nontrivial correlations; the signature of this is that the impurity self-energy vanishes with exactly the same power as the hybridization.\cite{gonzalez1998renormalization,fritz2004phase,logan2014common} Interestingly, we find within impurity-DMFT for PQD models that,
\begin{equation}
\label{eq:pseudogap_SE}
-\text{Im}~\Sigma_{\sigma}^{\text{loc}}(\omega)~\overset{|\omega|\ll T_K}{\sim}~ |\omega|^{\alpha}  \;,
\end{equation}
with the same exponent $\alpha$ as in Eq.~\ref{eq:pseudogap}. This is shown in the lower panel of Fig.~\ref{fig:hys_se}, and confirms that under self-consistency, impurity-DMFT generates a \emph{critical pseudogapped} effective single-impurity model. The resulting even orbital spectral function\cite{note:evenorb} then appears to give a very good approximation to the low-energy behavior arising in underscreened PQD models --- see Fig.~\ref{fig:comp_MLQD}.

The appearance of an effective pseudogapped local model for PQDs within impurity-DMFT is rather reminiscent of the results of Ref.~\onlinecite{PhysRevLett.97.096603,*PhysRevB.78.153304,*PhysRevLett.102.166806,*PhysRevB.87.205313}. In that work, a parallel double dot system was considered, with one interacting and one non-interacting dot. An exact local model was derived, integrating out the non-interacting conduction electron bath and also the other non-interacting dot; the effective hybridization was found to be pseudogapped with exponent $\alpha=2$. Our results therefore generalize these findings to fully interacting PQDs, albeit that our local models arise within the (approximate) impurity-DMFT scheme. Since impurity-DMFT is exact by construction for the case considered in Ref.~\onlinecite{PhysRevLett.97.096603,*PhysRevB.78.153304,*PhysRevLett.102.166806,*PhysRevB.87.205313}, it would be interesting explore this connection by tracking the evolution of the pseudogap as interactions on the second dot are switched on.


\subsection{Magnetic field}
\label{sec:mag}

\begin{figure}[t]
\begin{center}
\includegraphics[width=80mm]{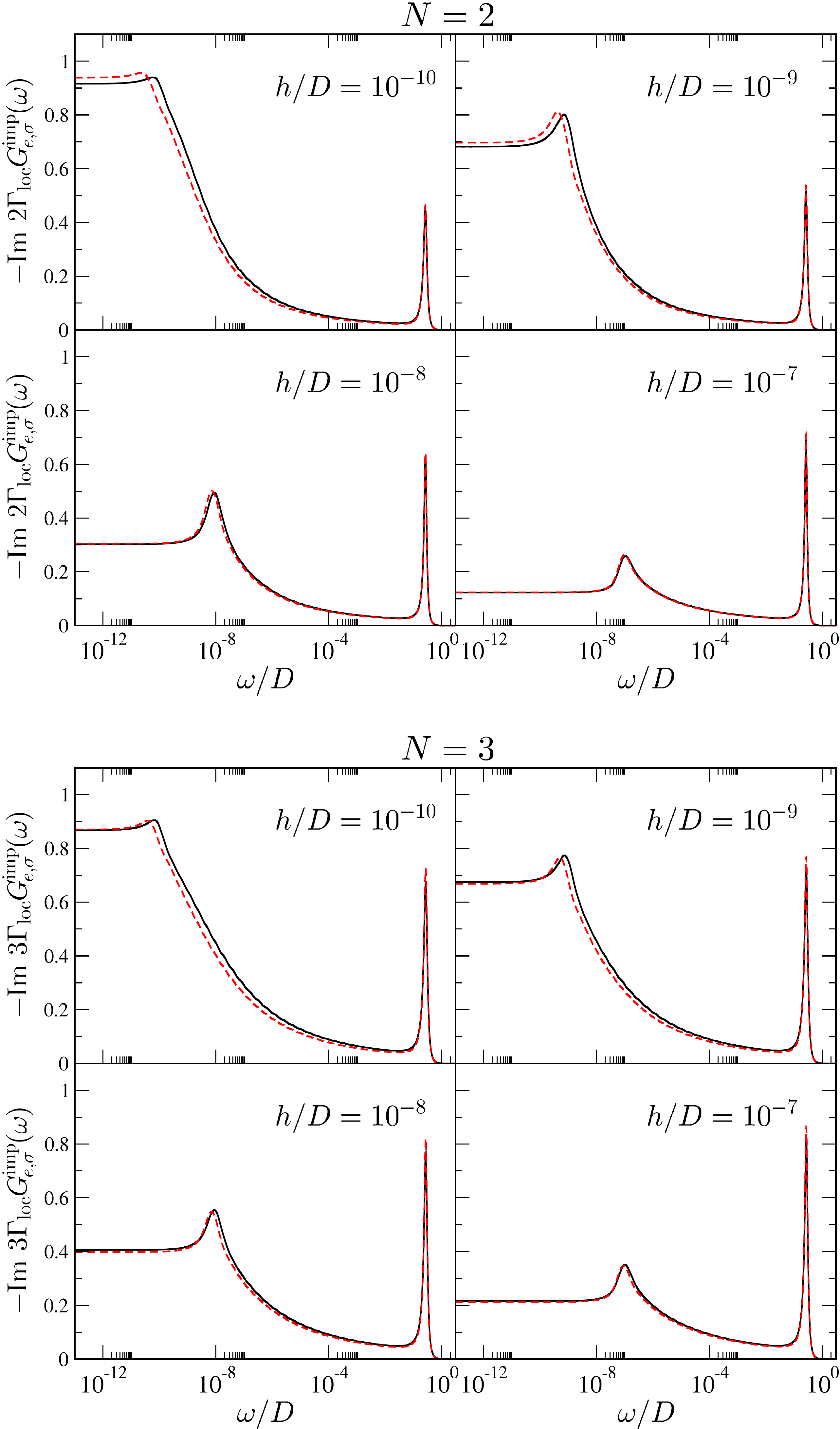}
\caption{\label{fig:comp_h}
Comparison of impurity-DMFT results (red dashed lines) and exact NRG (solid black lines) for the PQD  model with $N=2$ (upper panels) and $3$ (lower panels) in an applied magnetic field. Dot parameters $U/D=0.5$, $\epsilon=-U/2$ and $V/D=0.06$ such that $T^{h=0}_K\sim 10^{-9}D$. Shown for $v=0$ but $h/D=10^{-10}, 10^{-9}, 10^{-8}$ and $10^{-7}$. Plotted for $\sigma=\uparrow$.}
\end{center}
\end{figure}

\begin{figure}[t]
\begin{center}
\includegraphics[width=70mm]{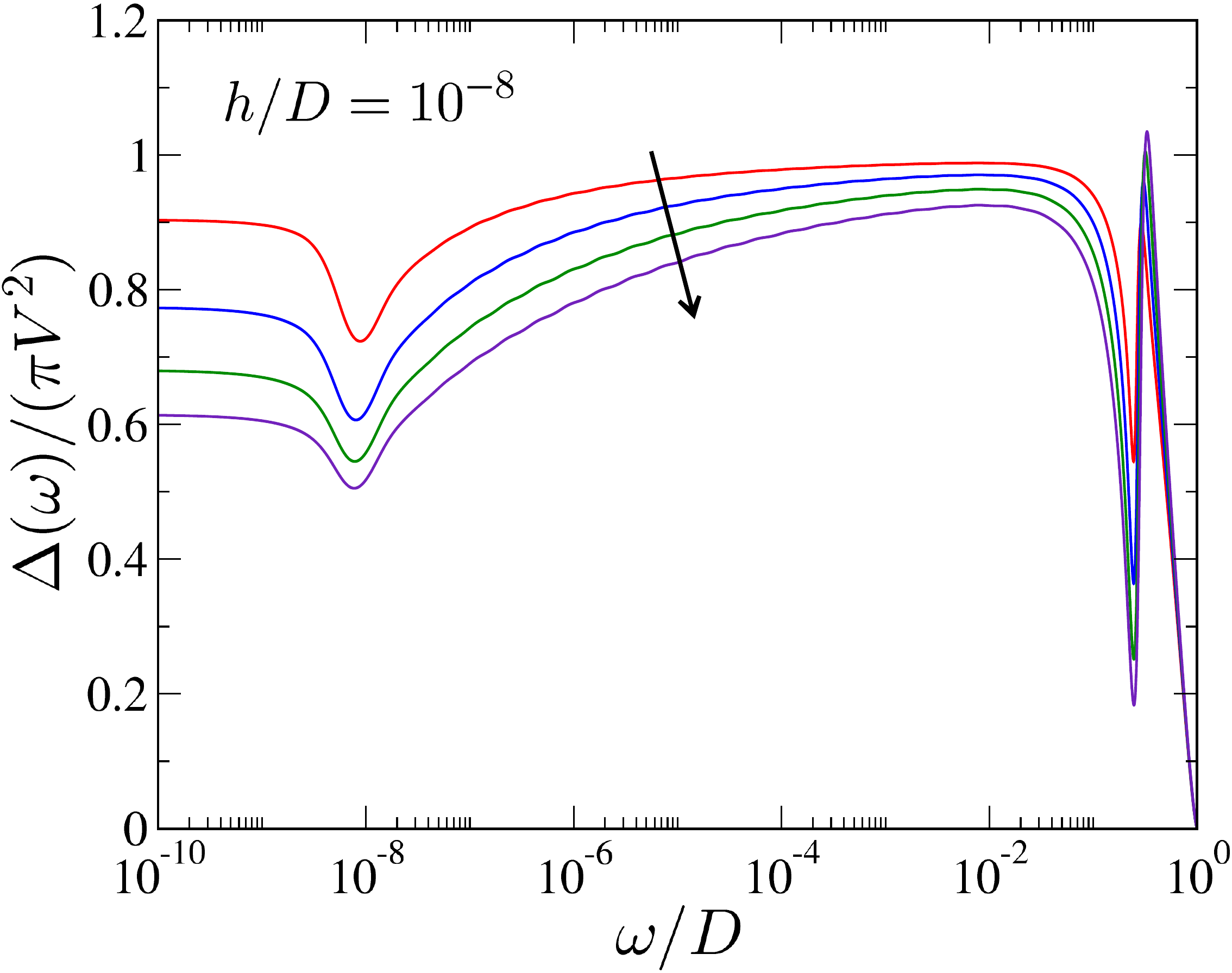}
\caption{\label{fig:hyb_h}
DMFT effective medium $\Delta(\omega)/(\pi V^2)$ vs $\omega/D$ for PQDs with $N=2,3,4$ and $5$ (increasing with the arrow) for $U/D=0.5$, $\epsilon=-U/2$, $V/D=0.06$ and $h/D=10^{-8}$ ($v=0$). }
\end{center}
\end{figure}

We consider now the effect of applying a magnetic field $h$ to the dots. The `impurity' Hamiltonian Eq.~\ref{eq:Himp} is supplemented by the term $H_{\text{mag}}= -h \hat{S}_{\text{PQD}}^z$, where $\hat{S}_{\text{PQD}}^z=\tfrac{1}{2}\sum_{\alpha}[\cre{d}{\alpha\uparrow}\ann{d}{\alpha\uparrow}-\cre{d}{\alpha\downarrow}\ann{d}{\alpha\downarrow}]$ is the \emph{total} PQD spin projection. This Zeeman term can be straightforwardly incorporated into the existing formalism by simply replacing $\epsilon \rightarrow \epsilon_{\sigma}$ in Eqs.~\ref{eq:CQD_dyson_e} and \ref{eq:AIM_G}, where $\epsilon_{\sigma}=\epsilon-h\sigma/2$ (and $\sigma=\pm$ for $\uparrow/\downarrow$).

In the standard single-impurity case $N=1$, an infinitesimal field $h$ does not destroy the Kondo singlet (which has a binding energy $\sim T_K$), and the impurity remains unmagnetized. On the other hand, for $h\gg T_K$ spin-flip scattering, which is at the heart of the Kondo effect, is suppressed.\cite{hewson1997kondo} 

In the present context of PQDs with $N>1$, the physics is more subtle: even for infinitesimal $h$, the residual dot spin-$(N-1)/2$, resulting from Kondo underscreening, becomes polarized. In terms of the dot spectral function, this implies a significant (spin-dependent) redistribution of spectral weight between the Hubbard satellites --- an effect \emph{not} observed in the single-impurity case. The same basic physics has been discussed for multi-level dots in Ref.~\onlinecite{wright2011magnetic}. For larger $h\gg T_K$, the Kondo effect is again destroyed. Most interesting is the case where the applied field $h\sim T_K$, so that the Kondo effect competes with spin polarization. Non-trivial physics associated with partial Kondo screening is then expected.

In Fig.~\ref{fig:comp_h} we present exact NRG results (solid black lines) for PQDs with $N=2$ (upper panels) and $N=3$ (lower panels), in the case of finite field $h/T_K \sim 0.1, 1, 10, 100$. This is the intermediate regime where the Kondo effect is only partially manifest due to competing spin polarization. In all cases, spectral weight is shifted between Hubbard satellites, indicative of dot magnetization. As $h$ increases, the Kondo resonance is progressively diminished, and a field-induced peak develops at $\omega\sim h$. The dot field cuts off the logarithmic corrections, and regularizes the spectrum for $|\omega| \ll h$.

The impurity-DMFT results are shown in Fig.~\ref{fig:comp_h} as the red dashed lines, and compare extremely well. Several nontrivial features of the full $N$-dot model are recovered by impurity-DMFT.

In particular, the redistribution of spectral weight to the upper Hubbard satellite is captured essentially exactly. This might seem surprising, because impurity-DMFT involves the mapping to a single-impurity model in which small fields are not expected to produce a large magnetization. Instead, the effect emerges from self-consistent development of structures in the effective medium $\Delta(\omega)$ on the scale of $|\omega|\sim U$, as shown in Fig.~\ref{fig:hyb_h}. The true PQD magnetization is in fact captured very well in impurity-DMFT.

The Fermi level value of the spectrum in Fig.~\ref{fig:comp_h} is also recovered with high accuracy by impurity-DMFT. This is a nontrivial result of the renormalized effective medium at the Fermi level, $\Delta(\omega=0)$  --- see Fig.~\ref{fig:hyb_h}. 

In addition, the spectral peak at $\omega \sim h$ (arising in these underscreened PQD models due to the asymmetric spin-resolved Kondo resonance\cite{wright2011magnetic}) is described very well by impurity-DMFT. Fig.~\ref{fig:hyb_h} again shows that this is a result of the self-consistent feedback of the local self-energy into the effective medium.

Indeed, the entire energy dependence of the spectrum is very well described by impurity-DMFT for both $N=2$ and $N=3$ over the full range of fields considered. The quality of the results improve with increasing $h$.


\subsection{Interdot coupling and quantum phase transition}
\label{sec:t}

\begin{figure}[t]
\begin{center}
\includegraphics[width=85mm]{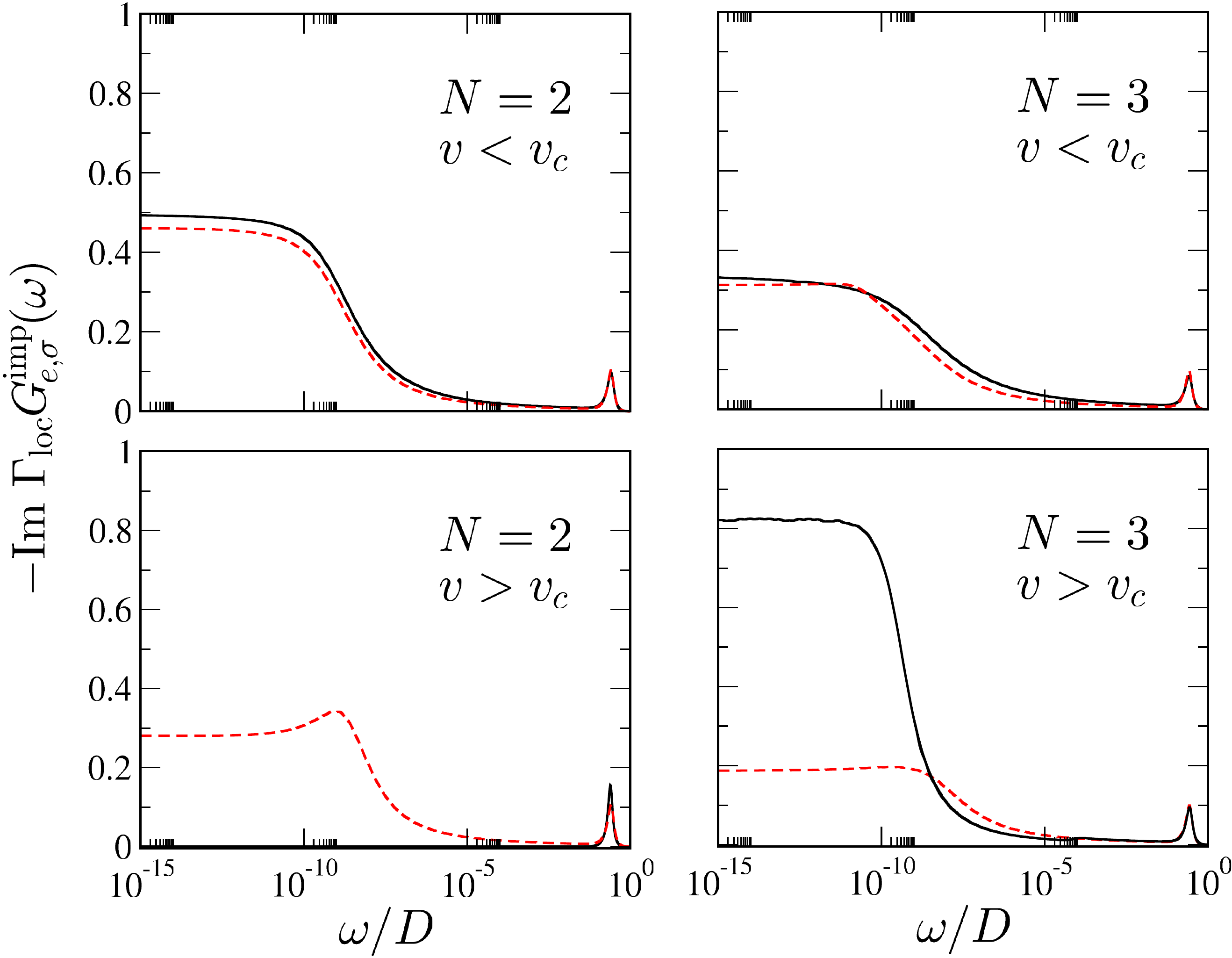}
\caption{\label{fig:comp_t}
Comparison of impurity-DMFT results (red dashed lines) and exact NRG (solid black lines) for the PQD  model with $N=2$ (left panels) and $3$ (right panels) with finite interdot coupling $v\ne 0$ across the quantum phase transition. Upper panels: high-spin phase with $v=0.005D<v_c$. Lower panels: low-spin phase with $v=0.01D>v_c$. Shown for fixed $U/D=0.5$, $\epsilon=-U/2$ and $V/D=0.06$ (but $h=0$).}
\end{center}
\end{figure}

Finally we consider the situation where the dots are also directly tunnel-coupled, $v\ne 0$. This generates a direct inter-dot exchange $J_{\text{dir}} \sim v^2/U$ which is \emph{antiferromagnetic}. Together with the lead-mediated RKKY coupling, the dots feel a mutual exchange $J_{\text{tot}}=J_{\text{RKKY}}+J_{\text{dir}}$ which can be tuned from ferromagnetic to antiferromagnetic by increasing the interdot tunnel-coupling $v$. The result of this is that a high-spin PQD state is realized for $v<v_c$ (corresponding $J_{\text{tot}}<0$), while a low-spin PQD state forms for $v>v_c$ ($J_{\text{tot}}>0$). These distinct phases are separated by a (level-crossing) quantum phase transition, arising at $v_c$ for which $J_{\text{tot}}=0$.

Interestingly, the Kondo physics for even and odd numbers of dots, $N$, is therefore very different, as discussed below. In Fig.~\ref{fig:comp_t} we consider specifically $N=2$ (left panels) and $N=3$ (right panels), in both the high-spin phase $v<v_c$ (upper panels) and the low-spin phase $v>v_c$ (lower panels). Exact NRG results for the $T=0$ spectral function are presented as the solid black lines.

In the high-spin phase $v<v_c$ (encompassing also the case $v=0$ discussed in Sec.~\ref{sec:PQDnrg}) the effective inter-dot ferromagnetic coupling $J_{\text{tot}}<0$ leads to a combined PQD dot spin-$N/2$ which is Kondo underscreened to spin-$(N-1)/2$ below $T_K$. This holds for even and odd $N$. The inclusion of finite $v$ in this phase does not change the spectral lineshape, which is universal for all $|\omega|\ll |J_{\text{tot}}|$ (only the value of $J_{\text{tot}}$ is changed). In particular, the low-energy spectral pinning condition Eq.~\ref{eq:pin} still holds. 

Including finite $v$ in the impurity-DMFT solution does not change the results qualitatively, although we note that the Fermi level value of the spectrum spuriously diminishes with increasing $v$ --- see upper panels of Fig.~\ref{fig:comp_t}.

By contrast, in the low-spin phase $v>v_c$ where $J_{\text{tot}}>0$ is antiferromagnetic, even and odd $N$ behave differently. For even $N$, the PQD forms a combined spin-singlet, and decouples from the leads on the scale of $|J_{\text{tot}}|$. As such, there is a total (on-the-spot) collapse of the Kondo resonance on entering the low-spin phase, 
$\text{Im}~\Gamma_{\text{loc}} G_{e,\sigma}^{\text{imp}}(\omega=0)=0$. This is seen from the exact NRG results for $N=2$ in the bottom-left panel of Fig.~\ref{fig:comp_t}. For odd $N$, the PQD forms an effective doublet state on the scale of $|J_{\text{tot}}|$, which is exactly screened by a standard Kondo effect below $T_K$. This leads to a pronounced spectral resonance, as seen for $N=3$ in the bottom right-panel of Fig.~\ref{fig:comp_t}. In this case
$-\text{Im}~\Gamma_{\text{loc}} G_{e,\sigma}^{\text{imp}}(0) < 1$ because the model is not particle-hole symmetric for finite $v$. An effective FSR\cite{hewson1997kondo} holds, $-\text{Im}~\Gamma_{\text{loc}} G_{e,\sigma}^{\text{imp}}(0)=\sin^2(\tfrac{\pi}{2} n_{\text{imp}}^e)$, where here the `excess charge'\cite{bulla2008numerical} of the PQD even orbital  $n_{\text{imp}}^e\ne 1$ depends on $v$.

Impurity-DMFT fails qualitatively to capture this quantum phase transition, as demonstrated by comparison to the red dashed lines. On increasing $v$, there is simply a smooth crossover, with $\text{Im}~\Gamma_{\text{loc}} G_{e,\sigma}^{\text{imp}}(\omega=0)\rightarrow 0$ as $v\rightarrow \infty$. In particular, the effective single-impurity Kondo physics for odd $N$ and $v>v_c$ is totally absent within impurity-DMFT. In this case, non-local self-energy contributions control the physics, and impurity-DMFT inevitably fails.


\section{Conclusion: applicability of inhomogeneous DMFT}
\label{sec:conc}

Although the self-energy becomes strictly local only in the limit of infinite dimensions,\cite{metzner1989correlated} DMFT employing local self-energies is widely used in the approximate treatment real systems with finite coordination.\cite{georges1996dynamical,held2007electronic}  Furthermore, inhomogeneous (or real-space) DMFT has been applied in diverse contexts, including e.g.\ layered materials\cite{freericks2004dynamical} and molecular electronics.\cite{florens2007nanoscale,jacob2010dynamical,turkowski2012dynamical}

DMFT is certainly a very powerful tool, providing deep insights into the physics of complex systems.\cite{georges1996dynamical} However, for highly inhomogeneous systems where both strong interactions and details of real-space geometry are important, it is often not obvious whether a local self-energy approximation could yield physically sensible results. Of course, the problems to which inhomogeneous DMFT is applied are typically those for which exact results are not available, and so the quality of the approximation can remain unclear.

In this paper we have applied ``impurity-DMFT'' to the 2IAM (featuring two magnetic impurities separated in real-space on the surface of a 3D cubic lattice), and PQD models (in which quantum dots are coupled in parallel between source and drain metallic leads). The local self-energy, combined with self-consistency, constitutes a rather sophisticated approximation, and nontrivial results are obtained.

Although the 2IAM and PQD models are among the simplest examples for which DMFT can be used, their exact solution with NRG is still very challenging.\cite{bulla2008numerical} By comparing exact NRG and impurity-DMFT results for dynamical quantities in these models, we were able to test the validity of the local self-energy approximation in highly inhomogeneous contexts.

We find that in situations where Kondo screening is the dominant effect, impurity-DMFT can work surprisingly well, especially at low energies. For example, in the 2IAM, impurity-DMFT correctly predicts the onset of the dilute limit,\cite{mitchell2015multiple} and recovers the Fermi level value of the impurity spectral function. In PQD models, subtle underscreened Kondo physics is apparently captured in impurity-DMFT through the self-consistent generation of a \emph{critical pseudogapped} effective medium. Impurity-DMFT is also correctly able to describe the effects of an applied magnetic field in PQD systems.

However, we also uncover cases where the local self-energy approximation fails catastrophically. These are situations where inter-impurity correlations are responsible for suppressing Kondo physics. For example, effective antiferromagnetic RKKY interactions in the 2IAM can lead to inter-impurity singlet states which simply cannot be described within impurity-DMFT. In PQD models, the quantum phase transition between high- and low-spin states on tuning inter-dot coupling is similarly not reproducible in impurity-DMFT.
These results naturally have implications for the wider applicability of DMFT to inhomogeneous systems.

Finally, we comment that exactly-solvable quantum impurity problems provide an ideal context to develop and test novel extensions of DMFT, going beyond the local self-energy approximation. \emph{Reliable} solutions will then be within reach for a wider range of problem.


\acknowledgements
A.K.M.~acknowledges funding from the D-ITP consortium, a program of the Netherlands Organisation for Scientific Research (NWO) that is funded by the Dutch Ministry of Education, Culture and Science (OCW). 
R.B.~acknowledges funding through the Institutional Strategy of the University of Cologne within the German Excellence Initiative. 
We are grateful for use of HPC resources at the University of Cologne.



\begin{thebibliography}{57}
\bibitem [{not()}]{note:evenorb}%
  \BibitemOpen
  \bibinfo {note} {The local and non-local dot Green functions
  $G_{\alpha\beta,\sigma}^{\text{imp}}(\omega)$ generated with impurity-DMFT
  feature non-physical divergences at the Fermi energy. However, these cancel
  in the even orbital Green function $G_{e,\sigma}^{\text{imp}}(\omega)$. The
  impurity-DMFT is self-consistent with respect to the even orbital Green
  function, because only the even orbital couples directly to the electronic
  host, Eq.~\ref{eq:Hhyb_CQD_e}.}\BibitemShut {Stop}%
\end{thebibliography}
%


\end{document}